\documentclass[sigconf,natbib=true]{acmart}
\AtBeginDocument{%
  }
\usepackage{multirow}
\usepackage{graphicx}
\usepackage{subcaption}
\usepackage{placeins}
\usepackage{caption}

\setcopyright{none}
\renewcommand\footnotetextcopyrightpermission[1]{}

\begin{document}

\title[Scaling Laws for Cross-Encoder Reranking]{Scaling Laws for Cross-Encoder Reranking}






\author{Rahul Seetharaman}
\affiliation{%
  \institution{UMass Amherst}
    \city{Amherst}
  \country{USA}
  }
\email{rseetharaman@umass.edu}

\author{Aman Bansal}
\affiliation{%
  \institution{UMass Amherst}
  \city{Amherst}
  \country{USA}
  }
\email{amanbansal@umass.edu}

\author{Hamed Zamani}
\affiliation{%
  \institution{UMass Amherst}
  \city{Amherst}
  \country{USA}
  }
\email{zamani@cs.umass.edu}

\author{Kaustubh D. Dhole}
\affiliation{%
  \institution{Emory University}
  \city{Atlanta}
  \country{USA}
}
\email{kdhole@emory.edu}

\begin{abstract}
Scaling laws are well studied for language models and first-stage retrieval, but not for reranking. We present the first systematic study of scaling laws for cross-encoder rerankers across pointwise, pairwise, and listwise objectives. Across model size and training exposure, ranking quality follows predictable power laws, enabling larger rerankers to be forecast from smaller runs. Using models up to 150M parameters, we forecast 400M and 1B rerankers on MSMARCO-dev and TREC DL. Beyond forecasting, we derive compute-allocation rules from the fitted joint scaling law and compare them with equal-compute checkpoints, showing that retrieval metrics often favor data-heavy scaling, though the recommendation depends on the training objective. The forecasts are accurate and typically conservative, making them useful for planning expensive large-model training. These results provide practical scaling principles for industrial reranking systems, and we will release code and evaluation protocols.
\end{abstract}

\newcommand{\kd}[1]{{\color{brown}[KD: #1]}} 
\newcommand{\hamed}[1]{{\color{blue}[Hamed: #1]}} 
\maketitle
\thispagestyle{plain}
\pagestyle{plain}

\section{Introduction}
\label{sec:intro}
Modern search engines commonly use multi-stage pipelines: a fast retriever such as BM25~\cite{robertson1995okapi} produces candidates, and a reranker refines their order~\cite{nogueira2019passage,wang2011cascade,hu2019retrieve,hofstatter2021intra,twitterRecommendationAlgorithm2023, nagarLiuUberDeliverySearch2025, johnsonLinkedInJobSearch2025, bingImageSearchDeepLearning2018, vorotilovShugaepovInstagramExplore2023}. Because reranking is the final high-precision stage, its quality strongly affects what users see.

Scaling laws are well established for language models~\cite{kaplan2020scaling}, dense retrieval~\cite{fang2024scalinglawsdenseretrieval, zeng2025scalingsparsedense}, and embedding models~\cite{embedding-scale}, but not for rerankers. That gap is important: rerankers operate on retriever-induced candidate sets, optimize heterogeneous learning-to-rank objectives, and are evaluated with discontinuous top-$k$ metrics such as NDCG. It is therefore unclear whether scaling trends from language modeling or first-stage retrieval transfer to reranking.

Our goal is to forecast large-reranker performance from smaller training runs, reducing the need for expensive 1B+ experiments. More broadly, we want scaling laws that not only predict performance, but also help decide how a fixed training budget should be split between larger models and more training exposure. Throughout the paper, we refer to objectives as \textit{pointwise}, \textit{pairwise}, and \textit{listwise}, corresponding to BCE, RankNet, and ListNet. We use this terminology consistently below. We study five questions for these rerankers:
\begin{itemize}
\item \textbf{RQ1 (Model Scaling):} \textit{Can we predict the performance of large reranking models from smaller reranking models trained on the same data?}
\item \textbf{RQ2 (Data Scaling):} \textit{With model size fixed, can we forecast later-stage ranking quality from earlier checkpoints?}


\item \textbf{RQ3 (Compute Scaling):} \textit{Can a joint law over model size and training exposure predict ranking quality across the full scaling grid?}
\item \textbf{RQ4 (Compute-Optimal Allocation):} \textit{Given a fixed training compute budget, what allocation between model size and training exposure is optimal for reranking quality?}
\item \textbf{RQ5 (Objective Sensitivity):} \textit{Do scaling laws differ across pointwise, pairwise, and listwise fine-tuning objectives?}
\end{itemize}

We present the first systematic scaling study of rerankers. Using cross-encoder models of varying sizes~\cite{bert-og, nogueira2019passage} fine-tuned on 100K MSMARCO queries, we show that NDCG follows smooth power laws across model, data, and joint scaling. Using checkpoints only up to 150M parameters, we accurately forecast 400M and 1B rerankers on MSMARCO-dev and TREC DL~\cite{craswell2020overviewtrec2019deep, craswell2025overviewtrec2023deep, trec-dl-21, trec-dl-22, mackie2021dlhard}. We run the data-scaling analysis for all six model sizes, but use the 150M model as the representative main-text slice and move the full size-by-size results to the appendix. The fitted joint law also yields simple compute-allocation rules, and matched-compute comparisons show that data-heavy scaling is often beneficial for pointwise and pairwise training, but not uniformly for listwise training.

\section{Background}
\label{sec:background}

\paragraph{Multi-Stage Retrieval and Reranking.}
Modern search systems typically use a fast first-stage retriever followed by a more expressive reranker. Because rerankers act on a retriever-induced candidate set and are evaluated with discontinuous top-$k$ metrics such as NDCG@10, their scaling behavior need not match that of pretraining or first-stage retrieval.

\paragraph{Scaling Laws and Forecasting in Machine Learning.}
Scaling laws describe how performance changes with model size, data, and compute. They are well established in language modeling~\cite{kaplan2020scaling}, vision~\cite{zhai2022scaling}, multimodal learning~\cite{goyal2023scaling}, and related settings~\cite{cortes1993learning,10.5555/3600270.3602446,kim2025pretraininginfinitecompute}. Yet forecasting \emph{downstream} metrics is harder than forecasting training loss~\cite{xu2025unveiling, chen2024scaling}, which is especially relevant in retrieval.

\paragraph{Scaling Laws in Retrieval.}
Recent work has begun to characterize scaling in \emph{first-stage} retrieval, including retrieval-augmented datastores~\cite{shao2024scalingretrievalstore}, sparse and dense retrieval~\cite{zeng2025scalingsparsedense}, and generative retrieval~\cite{cai2025generativeretrieval}. These studies concern candidate selection, not fine-grained reranking over a fixed candidate set.

\paragraph{Reranking Objectives and the Missing Scaling Picture.}
Rerankers score a small candidate set with richer interaction models, often cross-encoders, and are trained with pointwise, pairwise, or listwise objectives~\cite{liu2009learning,burges2005learning,joachims2002optimizing,cao2007learning,nogueira2019passage, ni2021large, macavaney2021sledge}. Despite extensive work on reranker design and efficiency~\cite{hofstatter2021, nogueira2020}, we still lack a principled account of how reranking quality scales with model size and training exposure. We address that gap by forecasting downstream NDCG@10 from smaller-scale runs, with contrastive entropy as a secondary diagnostic.

\section{Reranking Paradigms} 
\label{sec:paradigms}

Given a query $q$ and candidate set $\mathcal{C}(q) = \{d_1, \dots, d_K\}$, a reranker is a scoring function $f_\theta(q, d_i) \mapsto s_i \in \mathbb{R}$ that induces a ranking $\pi(q) = \operatorname{argsort}_i s_i$. We focus on three paradigms, and report all evaluations \emph{per paradigm}:

\begin{itemize}
    \item \textbf{Pointwise.} Each example is $(q, d, y)$. We use binary cross-entropy.

    \item \textbf{Pairwise.} Each example is $(q, d^+, d^-)$. We optimize the RankNet objective \cite{ranknet}.

    \item \textbf{Listwise.} Each example is $(q, \mathbf{d}, \mathbf{y})$. We optimize the ListNet objective \cite{listnet}.
\end{itemize}

\section{A Framework for Reranker Scaling Laws}
\label{sec:methodology}

We analyze reranker scaling by fitting simple parametric laws and measuring held-out forecasting error.
\subsection{Training, Fitting, and Evaluation Protocol}
\label{sec:fit-protocol}
\begin{enumerate}
    \item Train model families across size, training exposure, and their joint scaling.
    \item Evaluate NDCG@10, contrastive entropy (CE) and other metrics like MAP, MRR, etc.
    \item Fit power-law functions.
    \item Hold out late checkpoints and measure forecasting error.
\end{enumerate}

\subsection{Evaluation Metrics}
NDCG@10~\cite{NDCG} is our primary forecasting target. Because it is discontinuous, we also analyze contrastive entropy (CE) as a smoother secondary diagnostic, following prior dense-retrieval work~\cite{fang2024scalinglawsdenseretrieval}.
\begin{equation*}
 CE = -\log \frac{\exp(s(q_i, p_i^+; \theta))}{\exp(s(q_i, p_i^+; \theta)) + \sum_j \exp(s(q_i, p_j^-; \theta))}
\end{equation*}
Here, $q_i$ is a query, $p_i^+$ is a relevant passage, and $p_{ij}^-$ are sampled negatives. We compute CE over BM25 top-100 candidates, by sampling 64 candidates. We report held-out RMSE, since forecasting error is more useful than goodness-of-fit alone.

\subsection{Scaling}
We study three axes—model size, training exposure, and their joint effect—using power laws, which fit best among the functional forms we tried.\footnote{We also tested exponential, logarithmic, and polynomial variants, but power laws were the most consistent and predictive.}
\subsubsection{Model Size Scaling}
To characterize improvements with capacity, we fit each IR metric $\mathcal{M}$ as a function of model size $M$ using a saturating power law:
\begin{equation}
\mathrm{\mathcal{M}}(M) \;=\; a \;-\; b\,M^{-c},
\end{equation}
where $a$ is the asymptote and $c$ captures diminishing returns with scale.
\subsubsection{Data Size Scaling}

To isolate training progress, we treat data scaling as \emph{training exposure} and fit performance as a function of step $S$:

\begin{equation}
\mathrm{\mathcal{M}}(S) \;=\; a \;-\; b\,S^{-c},
\end{equation}
where $a$ is the plateau and $c$ controls how quickly performance saturates.

\subsubsection{Joint Scaling}
For simultaneous gains from larger models and more training, we fit:
\begin{equation}
\mathrm{\mathcal{M}}(M,S) \;=\; a \;-\; b\,M^{-\alpha} \;-\; c\,S^{-\beta},
\end{equation}
where $\alpha$ and $\beta$ capture diminishing returns along each axis.

\section{Experiments}\label{sec:experiments}
We use the Ettin cross-encoder series~\cite{ettin} at 17M, 32M, 68M, 150M, 400M, and 1B parameters, fine-tuned on 100K MS MARCO passage-ranking queries drawn from the public MS MARCO v1.1 release\footnote{\url{https://huggingface.co/datasets/microsoft/ms_marco}}. On Hugging Face, this English release contains 102K total examples split into 82.3K train, 10K validation, and 9.65K test rows, with query, passage, answer, and query-type metadata. Pointwise models use batch size 128; pairwise and listwise models use 16 queries with 1 positive and 10 anchor negatives (effective batch size 160). All models train for one epoch with learning rate $2\times 10^{-5}$.

We rerank the BM25 top-100 passages on MSMARCO-dev and also evaluate on TREC DL '19--'23 and DL Hard. For model scaling we use the last checkpoint of each model. For data scaling we use checkpoints across a single epoch. We fit these data-scaling curves for all six model sizes; for readability, the main text uses the 150M model as a representative middle-scale slice, while Appendix Tables~\ref{tab:data_scaling_all_models_msmarco} and \ref{tab:data_scaling_all_models_trec} report all sizes. For joint scaling we hold out the last five checkpoints per model size and fit on the rest.

\subsection{Statistical evaluation protocol}
We fit marginal curves with non-linear least squares and use the subtractive asymptotic form above for joint scaling. For data scaling we fit on all but the last five checkpoints; for model scaling we fit on the last ten checkpoints from the 17M, 32M, 68M, and 150M models and forecast the last ten checkpoints of the 400M and 1B models. We report fit quality via $R^2$, adjusted $R^2$, and significance against a constant baseline, along with held-out MAE/RMSE. For data scaling we additionally report the 95\% bootstrap interval at the last held-out checkpoint. For the held-out 400M and 1B forecasts, we also report 95\% bootstrap confidence intervals.

\section{Experimental Results}\label{sec:results}
Across MSMARCO-dev and TREC DL, two findings are consistent: reranker quality follows predictable scaling laws, and larger models can be forecast from smaller ones.

\subsection{Statistical validation of the scaling laws}

  The statistical results support the visual trends. On MSMARCO-dev, NDCG@10 data-scaling fits are strong across all three objectives,
  and model scaling from 17M--150M is tighter still (Table~\ref{tab:ndcg_scaling_fit_errors_all}). The main-text data row uses the 150M model as a representative slice, while Appendix Tables~\ref{tab:data_scaling_all_models_msmarco} and \ref{tab:data_scaling_all_models_trec} show that the same qualitative behavior holds across all six model sizes.
                     
  At the final checkpoint, the 95\% bootstrap intervals (500 resamples) cover 10 of the 12 held-out joint-law forecasts---across three objectives, two model sizes, and NDCG@10 and
  MAP---with the two misses both being pointwise at 400M (Table~\ref{tab:appendix_ndcg_fit_stats}).
  
  Contrastive Entropy tells a different story: the RMSE and MAE values are fairly higher (Table~\ref{tab:ce_scaling_fit_errors_all}). Besides, pairwise CE scaling is not consistently monotone even within a single model size (Figure~\ref{fig:contrastive_entropy_model_scaling}). This reflects CE's sensitivity to score calibration and margin fluctuations, which can shift independently of ranking quality. NDCG can improve even when CE is noisy, so CE is best treated as a coarse diagnostic rather than a primary forecasting target.
  
  On TREC DL, NDCG and MAP remain broadly predictable, but MRR is slightly noisier and TREC DL '19 is the clearest exception (Table \ref{tab:scaling_fit_errors_merged_trec}; Figure \ref{fig:mrr_trec_dl_19}).

\subsection{Scaling laws for NDCG}

\subsubsection{Model Scaling}
The following results describe the scaling behavior observed with respect to model size. 

\begin{table}[t]
\centering
\caption{NDCG@10 forecasting errors on MSMARCO-dev. The data-scaling row reports the representative 150M model; Appendix Tables~\ref{tab:data_scaling_all_models_msmarco} and \ref{tab:data_scaling_all_models_trec} report all model sizes. Sig. reports fit significance; detailed 95\% CI and $R^2$/adjusted $R^2$ values are deferred to Appendix Table~\ref{tab:appendix_ndcg_fit_stats}.}
\fontsize{7.0}{8.0}\selectfont
\setlength{\tabcolsep}{1.5pt}
\renewcommand{\arraystretch}{1.1}
\resizebox{\columnwidth}{!}{%
\begin{tabular}{llccccc}

\toprule
Scale & Obj. & RMSE & MAE & Sig. & Coef. \\
\midrule
\multirow{3}{*}{400M}
  & Pointwise & 0.010 & 0.009 & *** & $(0.347,-1.140)$ \\
  & Pairwise  & 0.009 & 0.009 & *** & $(0.366,-1.307)$ \\
  & Listwise  & 0.016 & 0.015 & *** & $(0.351,-1.179)$ \\
\midrule
\multirow{3}{*}{1B}
  & Pointwise & 0.013 & 0.013 & *** & $(0.347,-1.140)$ \\
  & Pairwise  & 0.012 & 0.012 & *** & $(0.366,-1.307)$ \\
  & Listwise  & 0.024 & 0.024 & *** & $(0.351,-1.179)$ \\
\midrule
\multirow{3}{*}{Data (150M)}
  & Pointwise & 0.030 & 0.029 & *** & $(0.492,-0.339)$ \\
  & Pairwise  & 0.026 & 0.025 & *** & $(0.443,-0.494)$ \\
  & Listwise  & 0.016 & 0.013 & *** & $(0.362,-0.796)$ \\
\midrule
\multirow{3}{*}{Joint}
  & Pointwise & 0.026 & 0.021 & *** & $(0.744,-1.137,-0.155)$ \\
  & Pairwise  & 0.023 & 0.019 & *** & $(0.467,-1.230,-0.430)$ \\
  & Listwise  & 0.026 & 0.022 & *** & $(0.432,-1.082,-0.449)$ \\
\bottomrule
\end{tabular}}
\label{tab:ndcg_scaling_fit_errors_all}
\end{table}

\textit{Observations:} Figure~\ref{fig:NDCG_model_scaling} shows clear model-scaling trends. Fitting on checkpoints up to 150M yields accurate forecasts for both 400M and 1B, with pointwise and pairwise objectives giving the lowest errors and listwise remaining slightly noisier.


\subsubsection{Data Scaling}
\textit{Observations:} NDCG also scales smoothly with training exposure. For readability, the main-text view shows the 150M model as a representative midpoint, but we fit the same data-scaling law for every model size and report those results in Appendix Tables~\ref{tab:data_scaling_all_models_msmarco} and \ref{tab:data_scaling_all_models_trec}. Gains plateau near the end of one epoch; pointwise saturates earlier, while pairwise and listwise improve for longer and achieve lower held-out error.
\subsubsection{Joint Model-Data scaling}



Figure~\ref{fig:NDCG_joint_powerlaw} and Table~\ref{tab:ndcg_scaling_fit_errors_all} show that NDCG remains predictable under joint scaling, indicating that the same subtractive asymptotic joint law captures the interaction between model size and training progress reasonably well.

\subsubsection{Multiplicative joint scaling law}
As an additional ablation, we also fit a multiplicative joint law, $\text{metric}(N,D)=a+bN^cD^e$, using the same train--test split. It captures the broad trend, but underfits relative to the additive law: for retrieval metrics its training $R^2$ is typically only $0.85$--$0.89$, versus roughly $0.91$--$0.94$ for the additive surface. Extrapolation to 400M and 1B is also weaker, especially for the pointwise objective, where the multiplicative 1B NDCG@10 MAE rises to $0.0898$ (Appendix Table~\ref{tab:appendix_mult_joint_ndcg}). We therefore treat the multiplicative form as a useful robustness check, but retain the additive formulation as the main joint-scaling model.


\subsection{Scaling laws on Contrastive Entropy (CE)}

\subsubsection{Model Scaling for CE}

We also fit the same laws to CE.



\subsubsection{Data Scaling for CE}
At 150M, CE is markedly less regular than NDCG. As with NDCG, we use 150M as the representative main-text slice while fitting all six model sizes in the appendix. In particular, pairwise data scaling is not consistently monotonic, suggesting that CE is more sensitive to score calibration than downstream ranking metrics.



\subsubsection{Joint model, data scaling}

\begin{figure*}[t]
    \centering

    \begin{subfigure}[t]{0.32\textwidth}
        \centering
        \includegraphics[width=\linewidth]{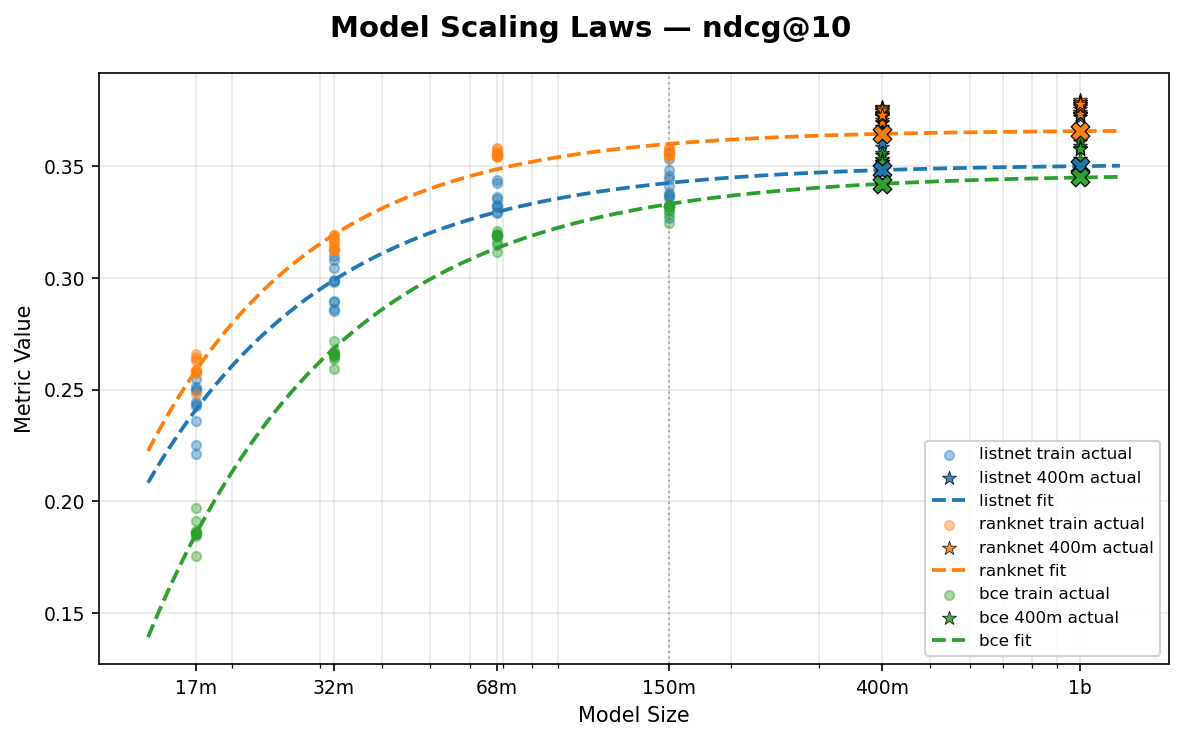}
        \caption{NDCG scaling with model size}
        \label{fig:NDCG_model_scaling}
    \end{subfigure}\hfill
    \begin{subfigure}[t]{0.32\textwidth}
        \centering
        \includegraphics[width=\linewidth]{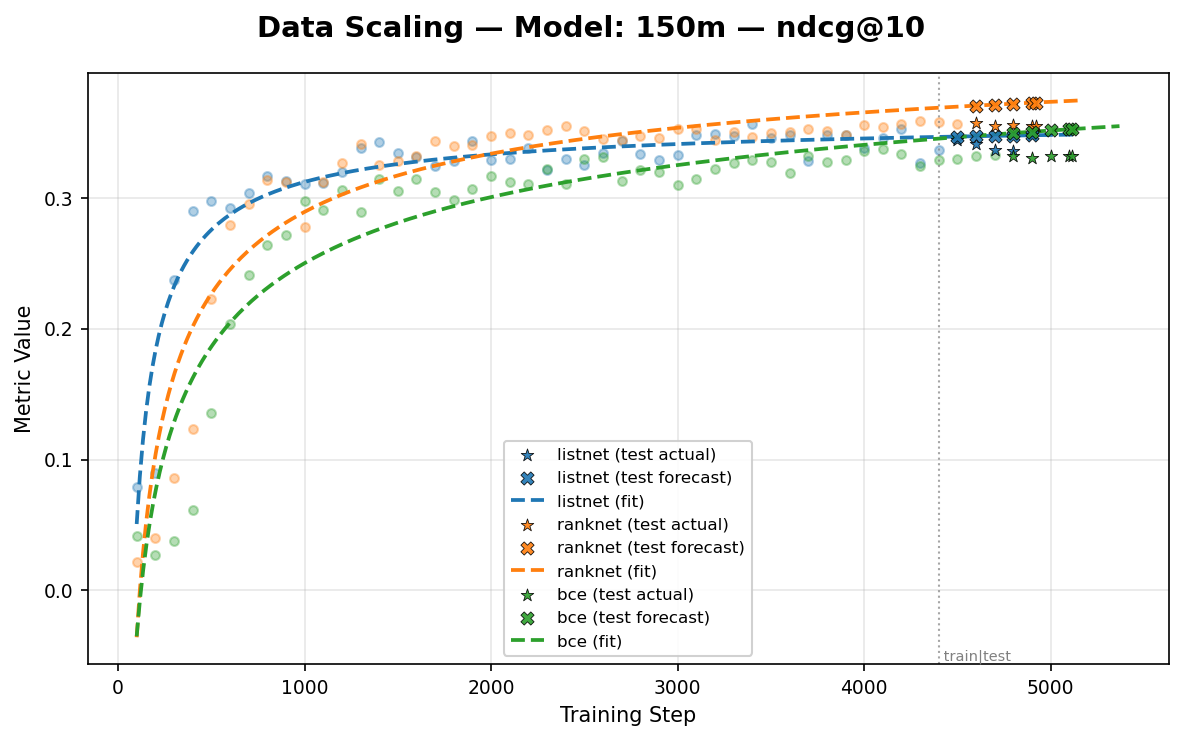}
        \caption{NDCG@10 data scaling for the 150M model}
        \label{fig:NDCG_data_scaling}
    \end{subfigure}\hfill
    \begin{subfigure}[t]{0.32\textwidth}
        \centering
        \includegraphics[width=\linewidth]{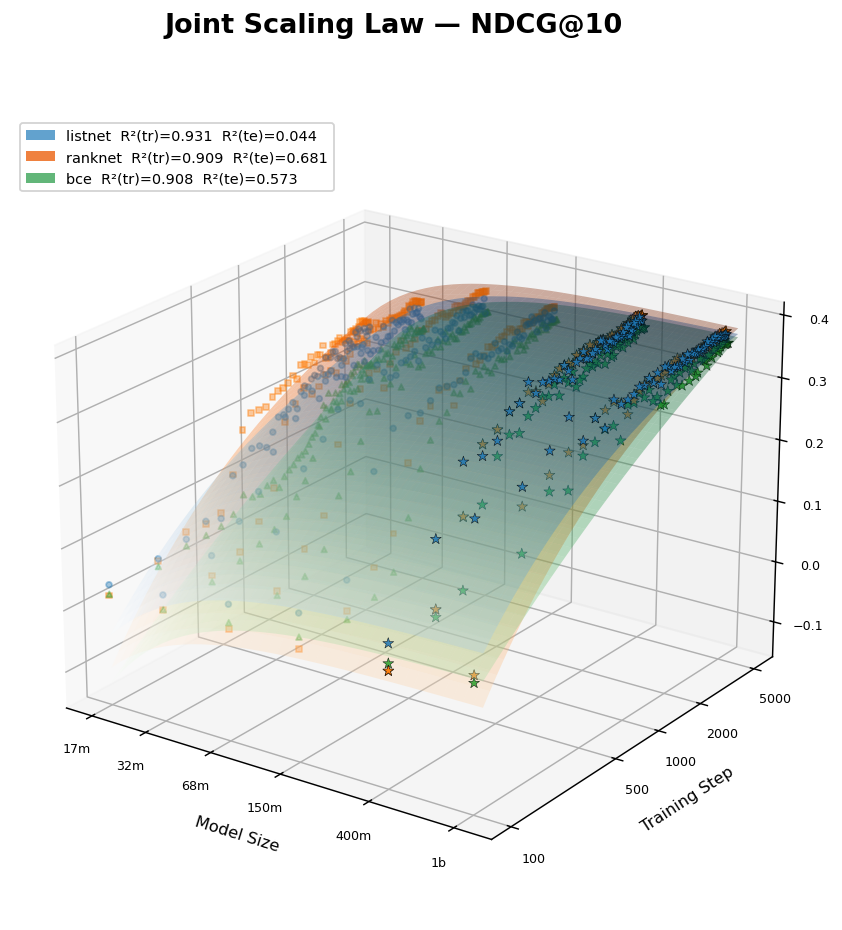}
        \caption{NDCG as a function of scaling with dataset and model sizes jointly}
        \label{fig:NDCG_joint_powerlaw}
    \end{subfigure}

    \vspace{0.6em}

    \begin{subfigure}[t]{0.32\textwidth}
        \centering
        \includegraphics[width=\linewidth]{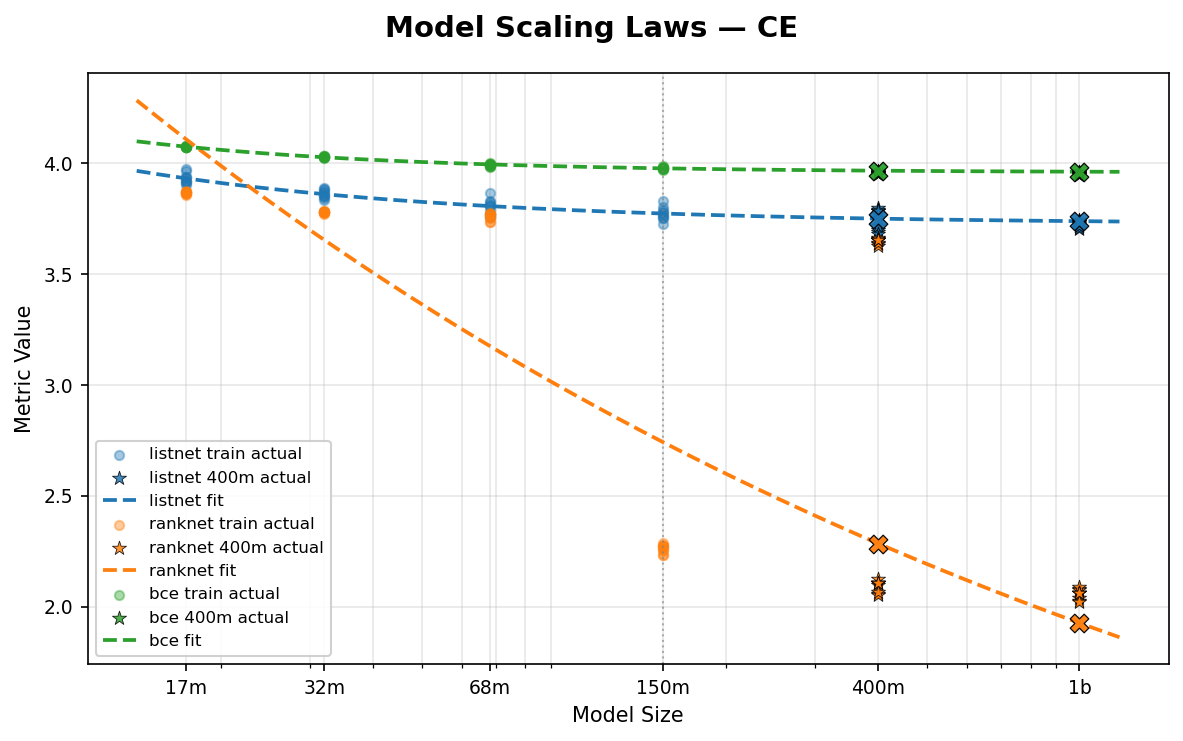}
        \caption{CE scaling with model size}
        \label{fig:contrastive_entropy_model_scaling}
    \end{subfigure}\hfill
    \begin{subfigure}[t]{0.32\textwidth}
        \centering
        \includegraphics[width=\linewidth]{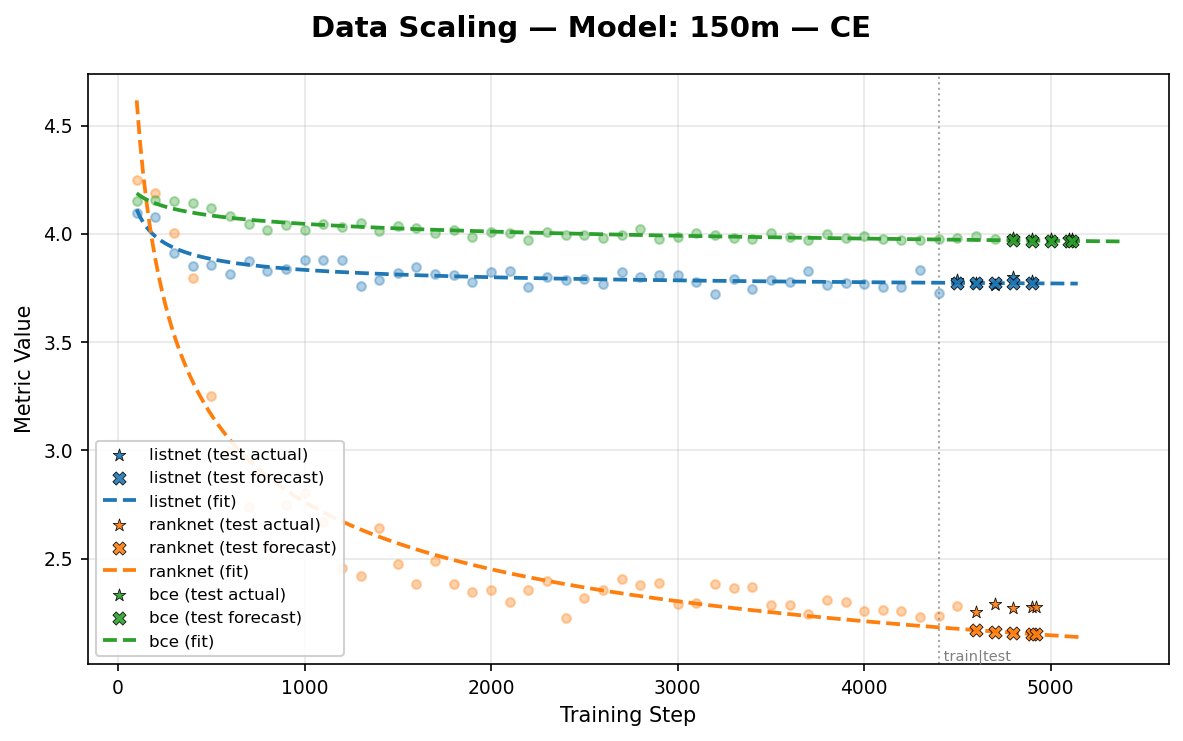}
        \caption{CE scaling with dataset size}
        \label{fig:contrastive_entropy_data_scaling}
    \end{subfigure}\hfill
    \begin{subfigure}[t]{0.32\textwidth}
        \centering
        \includegraphics[width=\linewidth]{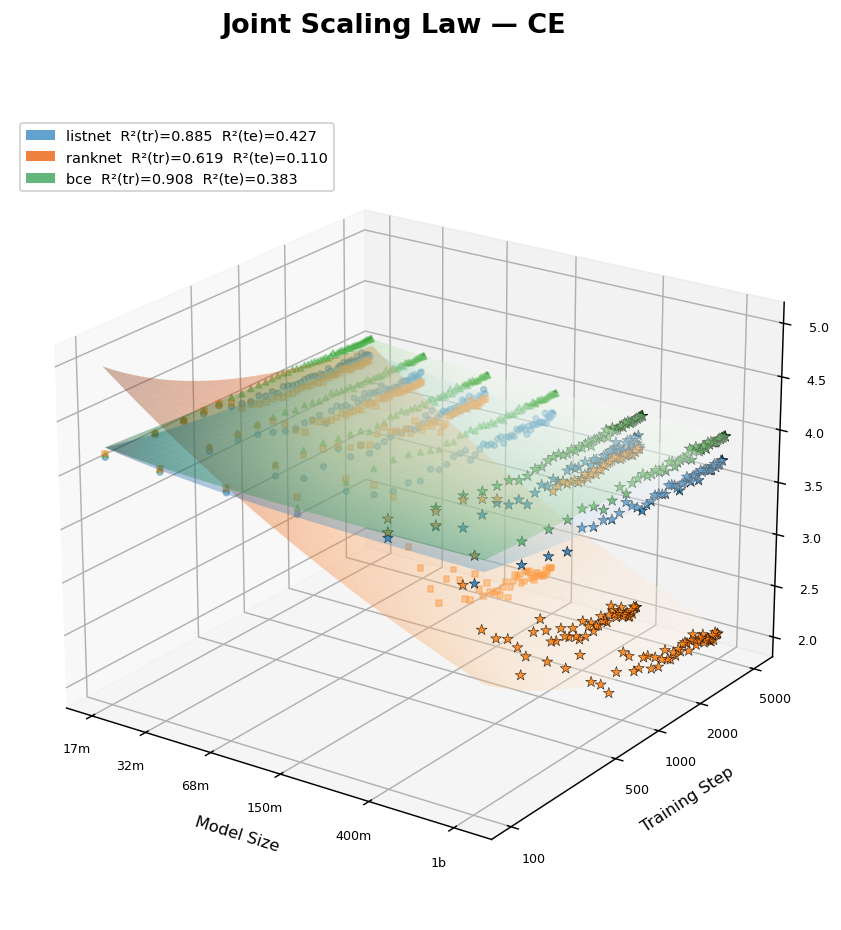}
        \caption{CE scaling with dataset and model sizes}
        \label{fig:contrastive_entropy_joint_model_data_scaling}
    \end{subfigure}

    \caption{Scaling behavior of NDCG@10 (panels a--c) and contrastive entropy (panels d--f) under model scaling, representative 150M-model data scaling, and joint scaling. Full data-scaling results for all six model sizes are reported in Appendix Tables~\ref{tab:data_scaling_all_models_msmarco} and \ref{tab:data_scaling_all_models_trec}.}
    \label{fig:scaling_grid_ndcg_ce}
\end{figure*}

Figure~\ref{fig:contrastive_entropy_joint_model_data_scaling} and Table~\ref{tab:ce_scaling_fit_errors_all} show the same pattern under joint scaling. Weak or inconsistent significance for CE is expected here and should be interpreted as evidence that CE is a less stable forecasting target than downstream retrieval metrics, rather than as a contradiction of our main claims. The multiplicative ablation does not resolve this instability: the pairwise CE case remains the weakest, with multiplicative RMSE $0.3817$ at 400M (Appendix Table~\ref{tab:appendix_mult_joint_ce}).



\begin{table}[t]
\centering
\caption{Contrastive entropy forecasting errors on MSMARCO-dev. The data-scaling row reports the representative 150M model; Appendix Tables~\ref{tab:data_scaling_all_models_msmarco} and \ref{tab:data_scaling_all_models_trec} report all model sizes. Sig. reports fit significance; detailed 95\% CI and $R^2$/adjusted $R^2$ values are deferred to Appendix Table~\ref{tab:appendix_ce_fit_stats}.}
\fontsize{7.0}{8.0}\selectfont
\setlength{\tabcolsep}{1.5pt}
\renewcommand{\arraystretch}{1.1}

\resizebox{\columnwidth}{!}{%
\begin{tabular}{llccccc}
\toprule
Scale & Obj. & RMSE & MAE & Sig. & Coef. \\
\midrule
\multirow{3}{*}{400M}
  & Pointwise & 0.011 & 0.010 & *** & $(3.959,-0.835)$ \\
  & Pairwise  & 0.975 & 0.778 & *** & $(0.000,-0.186)$ \\
  & Listwise  & 0.044 & 0.040 & *** & $(3.726,-0.673)$ \\
\midrule
\multirow{3}{*}{1B}
  & Pointwise & 0.010 & 0.009 & *** & $(3.959,-0.835)$ \\
  & Pairwise  & 0.133 & 0.131 & *** & $(0.000,-0.186)$ \\
  & Listwise  & 0.019 & 0.015 & *** & $(3.726,-0.673)$ \\
\midrule
\multirow{3}{*}{Data (150M)}
  & Pointwise & 0.129 & 0.124 & *** & $(3.604,-0.120)$ \\
  & Pairwise  & 0.131 & 0.128 & *** & $(1.395,-0.373)$ \\
  & Listwise  & 0.061 & 0.049 & *** & $(3.729,-0.563)$ \\
\midrule
\multirow{3}{*}{Joint}
  & Pointwise & 0.241 & 0.163 & *** & $(1.832,-0.876,-0.020)$ \\
  & Pairwise  & 0.153 & 0.123 & *** & $(0.000,-0.172,-0.459)$ \\
  & Listwise  & 0.105 & 0.089 & *** & $(3.599,-0.609,-0.348)$ \\
\bottomrule
\end{tabular}}

\label{tab:ce_scaling_fit_errors_all}
\end{table}


\subsection{Key findings}

\paragraph{\textbf{NDCG vs CE}} NDCG is easier to forecast than CE because ranking quality can improve even when score calibration and positive--negative margins fluctuate.

\paragraph{\textbf{Peak performance and model-size scaling.}}
  Across all retrieval metrics, the pairwise objective consistently achieves the highest absolute
  performance at large model sizes, reaching NDCG@10 of 0.378, Recall@10 of 0.549, and MAP of 0.329 at the 1B parameter scale---roughly 2 percentage points ahead of
  the listwise objective and 4--5 points ahead of the pointwise objective. This ordering is corroborated by the model-size scaling laws: the pairwise and pointwise objectives both fit power-law curves with high fidelity ($R^2 = 0.98$--$0.99$), while the listwise objective exhibits an anomalous drop in fit
  quality at 1B ($R^2 \approx 0.75$ for NDCG@10 and MAP), suggesting its training trajectory becomes less regular at large scale. The pointwise objective, despite trailing on absolute
  performance, produces the most predictable scaling behaviour---$R^2$ exceeding
  $0.99$ across every retrieval metric---making it the most reliable objective for extrapolating performance from smaller checkpoints.

  \paragraph{\textbf{Sensitivity to training data versus model size.}}
  The joint scaling analysis, fit to the form
  $\text{metric}(N, D) = a - b \cdot N^{c} - d \cdot D^{e}$,
  reveals a qualitative difference in how each objective responds to additional training
  compute. The listwise and pairwise objectives exhibit data exponents in the range
  $e \approx -0.43$ to $-0.50$, meaning performance continues to improve
  meaningfully as training steps increase. The pointwise objective's data exponent is substantially
  shallower ($e \approx -0.13$ to $-0.20$), indicating that its gains are driven
  almost entirely by model capacity rather than training duration. In practical
  terms, training a pointwise model longer yields diminishing returns far earlier than
  the listwise or pairwise objectives. This is consistent with the pointwise objective's higher theoretical asymptote
  ($a \approx 0.74$ for NDCG@10, versus $0.43$--$0.47$ for the other losses):
  the pointwise objective has greater headroom in principle, but requires disproportionately larger
  models---not more data---to approach it.

  \paragraph{\textbf{Cross-entropy instability under the pairwise objective.}}
  A notable exception to clean scaling behaviour occurs in the cross-entropy loss
  under the pairwise objective. While the listwise and pointwise objectives maintain smooth, monotonically improving CE
  trajectories across all model sizes (MAE $< 0.04$), the pairwise objective's CE degrades
  non-monotonically between 150M and 400M parameters
  (MAE $= 0.78$, $R^2 = 0.12$ at 400M), a breakdown that persists at 1B.
  This suggests that the pairwise ranking objective conflicts with
  likelihood calibration at larger capacities: the model increasingly sacrifices
  cross-entropy in favour of ranking order. Consequently, while the pairwise objective is the
  preferred choice when optimising retrieval rank metrics at scale, the listwise or
  pointwise objectives should be preferred in settings where cross-entropy calibration is also a
  requirement.

\subsection{Compute-Optimal Allocation Under the Joint Scaling Law}
\label{sec:compute_optimal}

We next ask how a fixed compute budget should be split between model size and training exposure. Because all models are trained for one epoch, training steps $D$ are proportional to unique training data consumed.

We define the training compute budget as
\begin{equation}
C = N \cdot D,
\end{equation}
where $N$ is the number of parameters and $D$ is the number of training steps; for fixed batch size and sequence length, this is proportional to training FLOPs.

We derive this frontier from the additive joint law rather than the multiplicative ablation. Under the multiplicative form $M(N,D)=a+bN^cD^e$, substituting $D=C/N$ gives $M(N;C)=a+bC^eN^{c-e}$, so the optimum lies on a boundary unless $c=e$; empirically, that variant also extrapolates less accurately on 400M and 1B.

Our fitted joint scaling law uses the same subtractive asymptotic form as Section~\ref{sec:methodology}:
\begin{equation}
M(N,D) = a - bN^{-\gamma} - dD^{-\delta},
\end{equation}
where $M$ denotes a retrieval metric or CE, and $(a,b,d,\gamma,\delta)$ are estimated from the joint fits over the 17M--150M models. This is algebraically equivalent to writing $a+bN^c+dD^e$ with $c=-\gamma$ and $e=-\delta$, but we use the subtractive form here to stay consistent with the rest of the paper and reserve $\alpha,\beta$ for the compute-optimal exponents below. Substituting $D=C/N$ yields
\begin{equation}
M(N;C) = a - bN^{-\gamma} - d\left(\frac{C}{N}\right)^{-\delta}
= a - bN^{-\gamma} - dC^{-\delta}N^{\delta}.
\end{equation}
Optimizing with respect to $N$ gives the closed-form allocation
\begin{align}
N^\star(C) &= \left(\frac{b\gamma}{d\delta}\right)^{\frac{1}{\gamma+\delta}} C^\alpha, \\
D^\star(C) &= \frac{C}{N^\star(C)}, \\
\alpha &= \frac{\delta}{\gamma+\delta}, \qquad \beta = \frac{\gamma}{\gamma+\delta}, \qquad \alpha+\beta=1.
\end{align}
Here, $\alpha$ governs optimal model scaling and $\beta$ optimal data scaling. When $\alpha<0.5$, the fitted law favors allocating more additional compute to data than to model size; when $\alpha>0.5$, it favors model scaling.

Table~\ref{tab:compute_exponents} reports these exponents. For all retrieval metrics, $\alpha<0.5$, implying a data-heavy allocation under the fitted law. The pointwise objective is the most data-favoring ($\alpha\approx0.10$--$0.15$), while the listwise and pairwise objectives are more moderate ($\alpha\approx0.25$--$0.37$). The only exception is pairwise CE ($\alpha=0.727$), reinforcing that CE behaves differently from downstream ranking metrics.

\begin{table}[t]
\centering
\scriptsize
\setlength{\tabcolsep}{3pt}
\resizebox{\columnwidth}{!}{%
\begin{tabular}{l l c c l}
\toprule
Objective & Metric & $\alpha$ ($N^\star \sim C^\alpha$) & $\beta$ ($D^\star \sim C^\beta$) & Regime \\
\midrule
Listwise & NDCG@10   & 0.293 & 0.707 & data-limited \\
Listwise & Recall@10 & 0.307 & 0.693 & data-limited \\
Listwise & MAP       & 0.286 & 0.714 & data-limited \\
Listwise & MRR       & 0.286 & 0.714 & data-limited \\
Listwise & CE        & 0.364 & 0.636 & data-limited \\
Listwise & P@10      & 0.306 & 0.694 & data-limited \\
\midrule
Pairwise & NDCG@10   & 0.259 & 0.741 & data-limited \\
Pairwise & Recall@10 & 0.280 & 0.720 & data-limited \\
Pairwise & MAP       & 0.249 & 0.751 & data-limited \\
Pairwise & MRR       & 0.250 & 0.750 & data-limited \\
Pairwise & CE        & 0.727 & 0.273 & model-limited \\
Pairwise & P@10      & 0.279 & 0.721 & data-limited \\
\midrule
Pointwise & NDCG@10   & 0.120 & 0.880 & data-limited \\
Pointwise & Recall@10 & 0.148 & 0.852 & data-limited \\
Pointwise & MAP       & 0.105 & 0.895 & data-limited \\
Pointwise & MRR       & 0.105 & 0.895 & data-limited \\
Pointwise & CE        & 0.022 & 0.978 & data-limited \\
Pointwise & P@10      & 0.148 & 0.852 & data-limited \\
\bottomrule
\end{tabular}
}
\caption{Compute-optimal scaling exponents derived analytically from the fitted joint scaling law. Values with $\alpha < 0.5$ indicate that the fitted law favors allocating more additional compute to data exposure than to model size.}
\label{tab:compute_exponents}
\end{table}

To place the observed runs relative to this frontier, we compute $N/N^\star(C)$ at each checkpoint. Values above $1$ indicate a model larger than the fitted optimum at the same compute budget. Table~\ref{tab:compute_deviation} shows that 400M and 1B are consistently above the frontier, whereas 17M and 32M are generally below it; 68M is closest under the listwise objective.

\begin{table}[t]
\centering
\scriptsize
\setlength{\tabcolsep}{3pt}
\resizebox{\columnwidth}{!}{%
\begin{tabular}{l l c c c l}
\toprule
Objective & Model & Median $N/N^\star$ & Min & Max & parameterization \\
\midrule
Listwise & 17M  & 0.376 & 0.309 & 1.010 & under \\
Listwise & 32M  & 0.592 & 0.487 & 1.545 & under \\
Listwise & 68M  & 1.012 & 0.833 & 2.569 & near \\
Listwise & 150M & 1.777 & 1.455 & 4.507 & over \\
Listwise & 400M & 3.537 & 2.890 & 9.048 & over \\
Listwise & 1B   & 6.678 & 5.484 & 17.352 & over \\
\midrule
Pairwise & 17M  & 0.530 & 0.449 & 1.208 & under \\
Pairwise & 32M  & 0.850 & 0.719 & 1.936 & under \\
Pairwise & 68M  & 1.492 & 1.251 & 3.396 & over \\
Pairwise & 150M & 2.661 & 2.229 & 6.126 & over \\
Pairwise & 400M & 5.363 & 4.561 & 12.729 & over \\
Pairwise & 1B   & 10.635 & 8.910 & 25.210 & over \\
\midrule
Pointwise & 17M  & 0.408 & 0.386 & 0.589 & under \\
Pointwise & 32M  & 0.730 & 0.669 & 1.033 & under \\
Pointwise & 68M  & 1.405 & 1.294 & 2.017 & over \\
Pointwise & 150M & 2.811 & 2.611 & 4.071 & over \\
Pointwise & 400M & 6.712 & 6.233 & 9.722 & over \\
Pointwise & 1B   & 15.135 & 14.054 & 22.309 & over \\
\bottomrule
\end{tabular}
}
\caption{Deviation of observed runs from the fitted compute-optimal frontier. $N/N^\star(C) > 1$ indicates that the observed model is larger than the model size preferred by the fitted joint law at the same compute budget.}
\label{tab:compute_deviation}
\end{table}

Because these exponents are fit-derived, we also compare observed checkpoints at matched compute. For each objective, we match the final 68M checkpoint to the 400M checkpoint with the same $C=N\cdot D$. Table~\ref{tab:equal_compute} summarizes these matched-compute comparisons.

The outcome is objective-dependent. Under the pairwise objective, 68M outperforms 400M at equal compute on all reported retrieval metrics. Under the pointwise objective, 68M wins on NDCG@10, MAP, and MRR, while Recall@10 and P@10 are nearly tied. Under the listwise objective, however, 400M remains better across all metrics.

\begin{table}[t]
\centering
\small
\setlength{\tabcolsep}{1.8pt}
\renewcommand{\arraystretch}{1.05}
\begin{tabular}{l p{1.12cm} r r r r r}
\toprule
Objective & Model & NDCG@10 & R@10 & MAP & MRR & P@10 \\
\midrule
Listwise & 68M fin. & 0.318 & 0.489 & 0.273 & 0.275 & 0.050 \\
Listwise & 400M eq. & \textbf{0.351} & \textbf{0.523} & \textbf{0.304} & \textbf{0.306} & \textbf{0.054} \\
\midrule
Pairwise & 68M fin. & \textbf{0.354} & \textbf{0.524} & \textbf{0.307} & \textbf{0.309} & \textbf{0.054} \\
Pairwise & 400M eq. & 0.342 & 0.516 & 0.294 & 0.297 & 0.053 \\
\midrule
Pointwise & 68M fin. & \textbf{0.319} & 0.483 & \textbf{0.275} & \textbf{0.277} & 0.050 \\
Pointwise & 400M eq. & 0.311 & \textbf{0.485} & 0.264 & 0.266 & \textbf{0.050} \\
\bottomrule
\end{tabular}
\caption{Equal-compute comparison using observed checkpoints. For each objective, we compare the 68M final checkpoint with the 400M checkpoint at matched compute budget ($C=N\cdot D$). Bold indicates the better value within each pair.}
\label{tab:equal_compute}
\end{table}

Taken together, the fitted exponents favor data-heavy scaling, but the direct equal-compute comparisons only partly confirm that picture. The pairwise and pointwise objectives show empirical evidence for data-limited behavior, whereas the listwise objective remains model-favored in the observed range. Compute-allocation recommendations for rerankers are therefore objective-dependent and should be validated with direct equal-compute comparisons.

\section{Evaluations on TREC DL Datasets}
We also evaluate on TREC DL '19--'23 and DL Hard~\cite{craswell2020overviewtrec2019deep,craswell2025overviewtrec2023deep, trec-dl-21, trec-dl-22, mackie2021dlhard}. NDCG shows predictable scaling on these benchmarks as well.




Table~\ref{tab:scaling_fit_errors_merged_trec} shows low prediction error for model, data, and joint scaling on TREC DL. As in the MSMARCO-dev analysis, the main-text data row uses the representative 150M model, while Appendix Table~\ref{tab:data_scaling_all_models_trec} reports all six model sizes. We leave broader cross-domain evaluation, e.g., on BEIR~\cite{beir}, to future work.

\begin{table}[t]
\centering
\caption{Model, data, and joint scaling-law fit errors (Test RMSE) on the TREC DL datasets across metrics. Each cell reports Test RMSE, with \texttt{***} indicating a statistically significant fit.}
\label{tab:scaling_fit_errors_merged_trec}
\fontsize{7.6}{8.6}\selectfont
\setlength{\tabcolsep}{2.4pt}
\renewcommand{\arraystretch}{1.15}

\begin{tabular*}{\columnwidth}{@{\extracolsep{\fill}} llrrr @{}}
\toprule
Scaling & Objective & NDCG & MAP & MRR \\
\midrule
\multirow{3}{*}{Model}
  & Pointwise & \shortstack{0.015\\\texttt{***}} & \shortstack{0.009\\\texttt{***}} & \shortstack{0.076\\\texttt{***}} \\
  & Pairwise  & \shortstack{0.032\\\texttt{***}} & \shortstack{0.007\\\texttt{***}} & \shortstack{0.053\\\texttt{***}} \\
  & Listwise  & \shortstack{0.035\\\texttt{***}} & \shortstack{0.007\\\texttt{***}} & \shortstack{0.053\\\texttt{***}} \\
\midrule
\multirow{3}{*}{Data (150M)}
  & Pointwise & \shortstack{0.018\\\texttt{***}} & \shortstack{0.005\\\texttt{***}} & \shortstack{0.027\\\texttt{***}} \\
  & Pairwise  & \shortstack{0.017\\\texttt{***}} & \shortstack{0.007\\\texttt{***}} & \shortstack{0.029\\\texttt{***}} \\
  & Listwise  & \shortstack{0.011\\\texttt{***}} & \shortstack{0.005\\\texttt{***}} & \shortstack{0.019\\\texttt{***}} \\
\midrule
\multirow{3}{*}{Joint}
  & Pointwise & \shortstack{0.028\\\texttt{***}} & \shortstack{0.042\\\texttt{***}} & \shortstack{0.103\\\texttt{***}} \\
  & Pairwise  & \shortstack{0.025\\\texttt{***}} & \shortstack{0.048\\\texttt{***}} & \shortstack{0.091\\\texttt{***}} \\
  & Listwise  & \shortstack{0.022\\\texttt{***}} & \shortstack{0.045\\\texttt{***}} & \shortstack{0.106\\\texttt{***}} \\
\bottomrule
\end{tabular*}
\end{table}

\section{Evaluations on other key IR metrics}


MAP is also consistently predictable, while MRR is more stable on MSMARCO-dev than on TREC. Across both metrics, pairwise and listwise objectives generally scale better than pointwise, though the exact trend varies by benchmark.


\begin{table}[t]
\centering
\caption{Scaling-law forecasting errors (Test RMSE) for MAP and MRR on MSMARCO-dev. For model and joint scaling, the laws are fit on models up to 150M and evaluated separately on 400M and 1B. \texttt{***} indicates a statistically significant fit.}
\label{tab:scaling_map_mrr_rmse}
\fontsize{7.6}{8.6}\selectfont
\setlength{\tabcolsep}{2.2pt}
\renewcommand{\arraystretch}{1.15}

\begin{tabular*}{\columnwidth}{@{\extracolsep{\fill}} llcccc @{}}
\toprule
Scale & Obj. & MAP & Sig. & MRR & Sig. \\
\midrule
\multirow{3}{*}{400M}
  & Pointwise & 0.010 & \texttt{***} & 0.010 & \texttt{***} \\
  & Pairwise  & 0.007 & \texttt{***} & 0.007 & \texttt{***} \\
  & Listwise  & 0.016 & \texttt{***} & 0.016 & \texttt{***} \\
\midrule
\multirow{3}{*}{1B}
  & Pointwise & 0.014 & \texttt{***} & 0.013 & \texttt{***} \\
  & Pairwise  & 0.010 & \texttt{***} & 0.010 & \texttt{***} \\
  & Listwise  & 0.022 & \texttt{***} & 0.023 & \texttt{***} \\
\midrule
\multirow{3}{*}{Data}
  & Pointwise & 0.025 & \texttt{***} & 0.025 & \texttt{***} \\
  & Pairwise  & 0.023 & \texttt{***} & 0.023 & \texttt{***} \\
  & Listwise  & 0.016 & \texttt{***} & 0.016 & \texttt{***} \\
\midrule
\multirow{3}{*}{Joint}
  & Pointwise & 0.021 & \texttt{***} & 0.021 & \texttt{***} \\
  & Pairwise  & 0.019 & \texttt{***} & 0.020 & \texttt{***} \\
  & Listwise  & 0.022 & \texttt{***} & 0.023 & \texttt{***} \\
\bottomrule
\end{tabular*}
\end{table}




\begin{figure*}[t]
    \centering

    \begin{subfigure}[t]{0.32\textwidth}
        \centering
        \includegraphics[width=\linewidth]{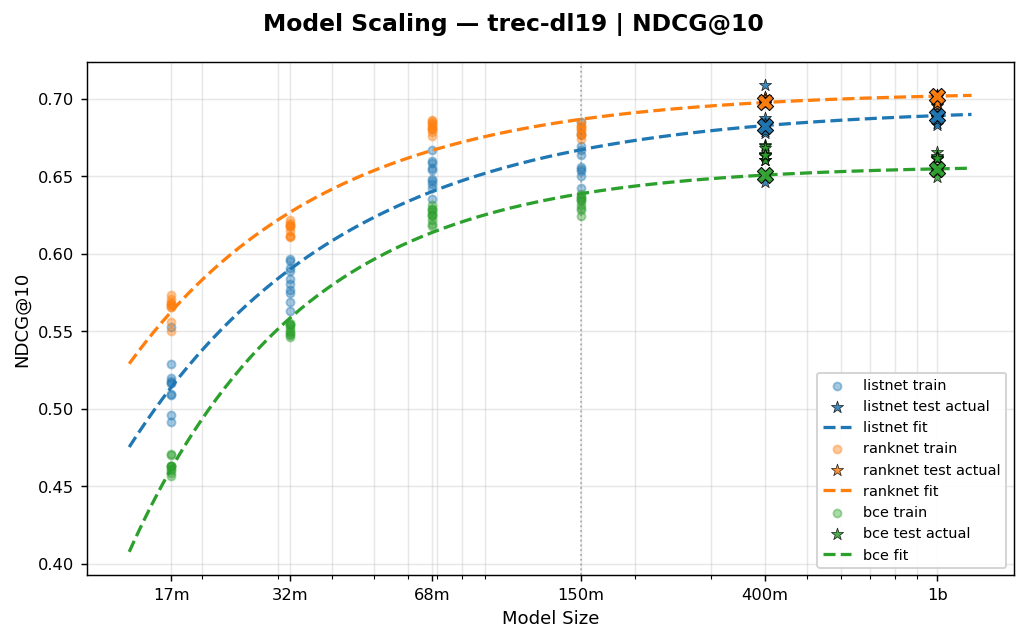}
        \caption{NDCG@10 on TREC DL '19.}
        \label{fig:ndcg_scaling_trec_dl19}
    \end{subfigure}\hfill
    \begin{subfigure}[t]{0.32\textwidth}
        \centering
        \includegraphics[width=\linewidth]{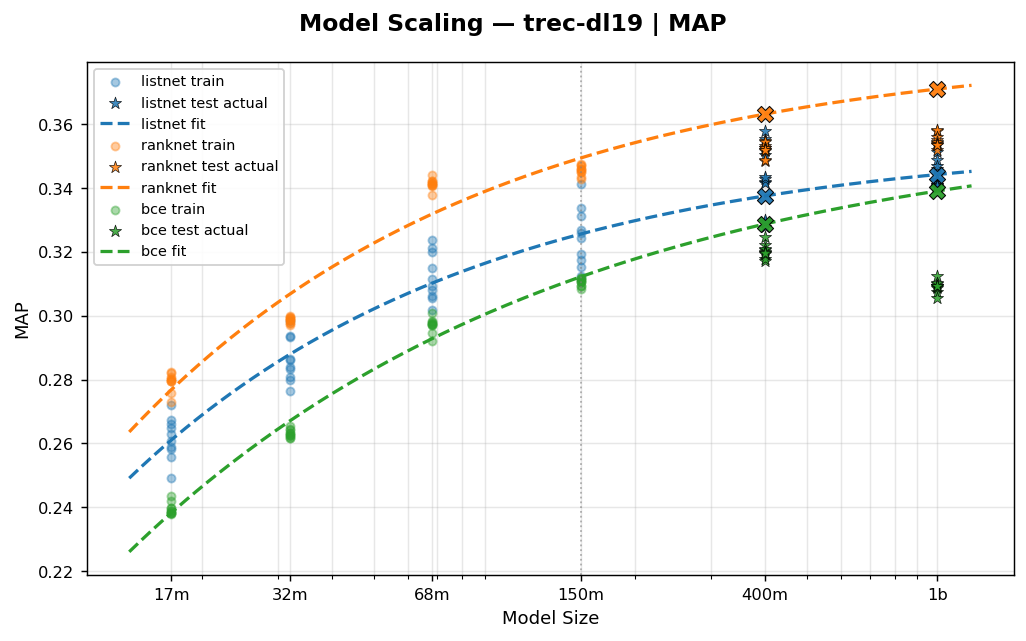}
        \caption{MAP on TREC DL '19.}
        \label{fig:map_model_scaling_trec_dl19}
    \end{subfigure}\hfill
    \begin{subfigure}[t]{0.32\textwidth}
        \centering
        \includegraphics[width=\linewidth]{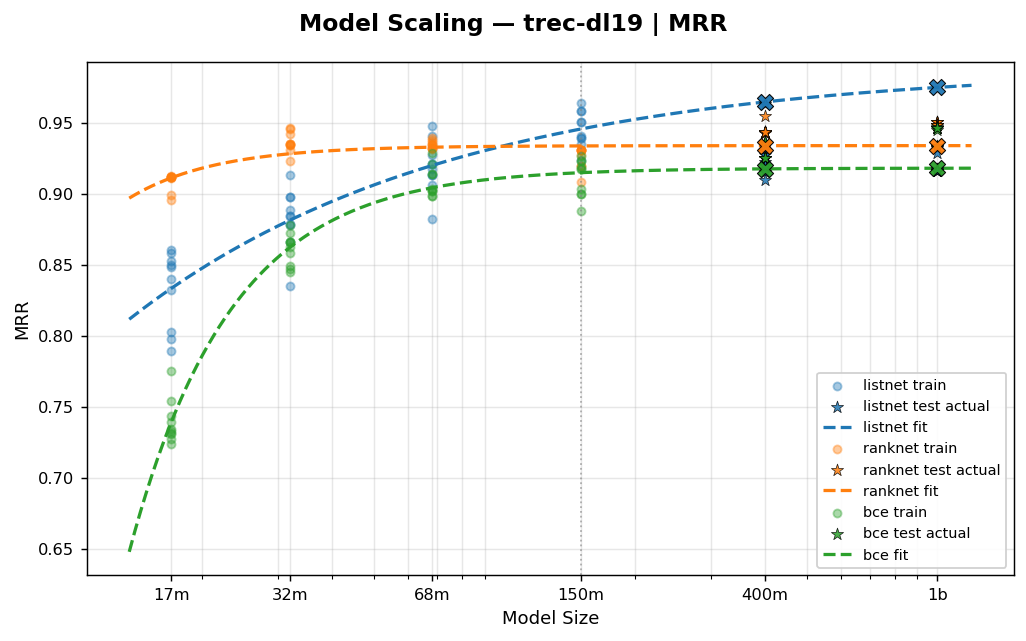}
        \caption{MRR on TREC DL '19.}
        \label{fig:mrr_trec_dl_19}
    \end{subfigure}

\caption{On the TREC DL '19 benchmark, model-scaling trends show that NDCG@10 and MAP scale predictably with model size, whereas MRR is less well behaved.}
    \label{fig:trec_dl19_scaling_1x3}
\end{figure*}

Finally, MRR shows usable scaling on MSMARCO-dev and some TREC sets, but is noticeably noisier than NDCG and MAP; TREC DL '19 is the clearest exception (Figure~\ref{fig:mrr_trec_dl_19}).

\section{Limitations}
\label{sec:limitations}
Our study is intentionally narrow in several ways. First, all experiments use a single encoder-based cross-encoder model family, so the fitted exponents may not transfer directly to other reranker families or architectures. Second, we evaluate on MSMARCO-dev and TREC DL, but do not report BEIR results, so our evidence for broader cross-domain generalization remains limited. Third, all models are fine-tuned for only one epoch on 100K MS MARCO queries. This keeps the comparison controlled across sizes and objectives, but it leaves open whether longer training horizons, larger supervision sets, or different curricula would change the observed scaling laws.

\section{Conclusion}
\label{sec:conclusion}
We present the first systematic study of scaling laws for reranking. Across pointwise, pairwise, and listwise rerankers, NDCG follows predictable power laws over model size and training exposure, allowing 400M and 1B performance to be forecast from models up to 150M. We run data scaling for all six model sizes and use 150M as the representative main-text slice, with the full size-by-size results reported in the appendix. MAP shows similar behavior, while MRR and CE are noisier and less reliable. Overall, small-scale sweeps can guide reranker scaling decisions, while objective choice remains an important source of variation.

\FloatBarrier

\bibliographystyle{ACM-Reference-Format}
\bibliography{sample-base}

\appendix
\section{Appendix}

Table~\ref{tab:bootstrap_ci_final_ckpt} reports the observed values, point forecasts, and 95\% bootstrap confidence intervals for the final-checkpoint joint-law predictions on the held-out 400M and 1B models. Tables~\ref{tab:appendix_ndcg_fit_stats} and \ref{tab:appendix_ce_fit_stats} move the detailed $R^2$/adjusted $R^2$ fit summaries for the main-text NDCG and CE tables to the appendix. Tables~\ref{tab:appendix_mult_joint_ndcg} and \ref{tab:appendix_mult_joint_ce} summarize the multiplicative joint-law ablation for NDCG@10 and CE. Tables~\ref{tab:data_scaling_all_models_msmarco} and \ref{tab:data_scaling_all_models_trec} report held-out data-scaling forecast errors for all model sizes.

Across MSMARCO-dev, the main trend from the 150M model extends cleanly to the full size range. Held-out errors remain small for all three objectives, and the curves are especially stable at larger model sizes. Listwise generally yields the lowest or near-lowest errors for NDCG@10 and MAP, while pointwise and pairwise remain competitive across the smaller models. MRR follows nearly the same pattern as MAP on MSMARCO-dev, suggesting that in-domain data scaling is broadly predictable once the reranker family and objective are fixed.

\begin{table*}[t]
\centering
\footnotesize
\caption{Observed values, point forecasts, and 95\% bootstrap confidence intervals for final-checkpoint joint-scaling predictions on MSMARCO-dev. Coverage is with respect to the observed held-out value.}
\label{tab:bootstrap_ci_final_ckpt}
\setlength{\tabcolsep}{3.5pt}
\renewcommand{\arraystretch}{1.15}
\begin{tabular*}{\textwidth}{@{\extracolsep{\fill}} llcccccc @{}}
\toprule
Metric & Objective & Obs. 400M & Pred. 400M & 95\% Boot CI (400M) & Obs. 1B & Pred. 1B & 95\% Boot CI (1B) \\
\midrule
\multirow{3}{*}{NDCG@10}
  & Listwise   & 0.3648 & 0.3704 & [0.3625, 0.3782] & 0.3733 & 0.3731 & [0.3634, 0.3829] \\
  & Pairwise   & 0.3730 & 0.3811 & [0.3725, 0.3897] & 0.3780 & 0.3816 & [0.3714, 0.3918] \\
  & Pointwise  & 0.3519 & 0.3742 & [0.3574, 0.3909] & 0.3576 & 0.3793 & [0.3563, 0.4023] \\
\midrule
\multirow{3}{*}{MAP}
  & Listwise   & 0.3178 & 0.3191 & [0.3131, 0.3250] & 0.3245 & 0.3236 & [0.3147, 0.3324] \\
  & Pairwise   & 0.3239 & 0.3293 & [0.3227, 0.3359] & 0.3297 & 0.3297 & [0.3226, 0.3367] \\
  & Pointwise  & 0.3050 & 0.3204 & [0.3052, 0.3356] & 0.3103 & 0.3270 & [0.3079, 0.3461] \\
\bottomrule
\end{tabular*}
\end{table*}

\begin{table*}[t]
\centering
\footnotesize
\caption{Fit diagnostics moved from the main-text NDCG@10 table. The CI column reports the 95\% bootstrap forecast interval for model and joint scaling, and the 95\% bootstrap interval at the last held-out checkpoint for data scaling across all six model sizes. Rows report training-set $R^2$/adjusted $R^2$, and Sig. gives the corresponding fit-significance code.}
\label{tab:appendix_ndcg_fit_stats}
\setlength{\tabcolsep}{3.2pt}
\renewcommand{\arraystretch}{1.1}
\begin{tabular*}{\textwidth}{@{\extracolsep{\fill}} llccc @{}}
\toprule
Scale & Objective & 95\% CI & Train $R^2$/Adj. & Sig. \\
\midrule
\multirow{3}{*}{Model}
  & Pointwise & \shortstack{$400M$: [0.339, 0.347]\\$1B$: [0.342, 0.350]} & 0.994/0.993 & *** \\
  & Pairwise  & \shortstack{$400M$: [0.362, 0.369]\\$1B$: [0.363, 0.371]} & 0.982/0.981 & *** \\
  & Listwise  & \shortstack{$400M$: [0.343, 0.356]\\$1B$: [0.344, 0.360]} & 0.953/0.950 & *** \\
\midrule
\multirow{3}{*}{Data (17M)}
  & Pointwise & [0.170, 0.195] & 0.824/0.816 & *** \\
  & Pairwise  & [0.257, 0.283] & 0.897/0.892 & *** \\
  & Listwise  & [0.244, 0.259] & 0.927/0.924 & *** \\
\midrule
\multirow{3}{*}{Data (32M)}
  & Pointwise & [0.277, 0.296] & 0.909/0.905 & *** \\
  & Pairwise  & [0.313, 0.343] & 0.901/0.896 & *** \\
  & Listwise  & [0.292, 0.314] & 0.921/0.917 & *** \\
\midrule
\multirow{3}{*}{Data (68M)}
  & Pointwise & [0.315, 0.348] & 0.874/0.868 & *** \\
  & Pairwise  & [0.352, 0.382] & 0.897/0.892 & *** \\
  & Listwise  & [0.328, 0.340] & 0.953/0.951 & *** \\
\midrule
\multirow{3}{*}{Data (150M)}
  & Pointwise & [0.327, 0.363] & 0.862/0.856 & *** \\
  & Pairwise  & [0.353, 0.383] & 0.898/0.893 & *** \\
  & Listwise  & [0.336, 0.355] & 0.894/0.889 & *** \\
\midrule
\multirow{3}{*}{Data (400M)}
  & Pointwise & [0.342, 0.369] & 0.894/0.890 & *** \\
  & Pairwise  & [0.367, 0.382] & 0.936/0.935 & *** \\
  & Listwise  & [0.356, 0.366] & 0.963/0.961 & *** \\
\midrule
\multirow{3}{*}{Data (1B)}
  & Pointwise & [0.348, 0.358] & 0.945/0.942 & *** \\
  & Pairwise  & [0.369, 0.378] & 0.974/0.972 & *** \\
  & Listwise  & [0.369, 0.378] & 0.758/0.747 & *** \\
\midrule
\multirow{3}{*}{Joint}
  & Pointwise & \shortstack{$400M$: [0.356, 0.393]\\$1B$: [0.358, 0.411]} & 0.908/0.906 & *** \\
  & Pairwise  & \shortstack{$400M$: [0.371, 0.390]\\$1B$: [0.372, 0.392]} & 0.909/0.907 & *** \\
  & Listwise  & \shortstack{$400M$: [0.362, 0.379]\\$1B$: [0.363, 0.384]} & 0.931/0.930 & *** \\
\bottomrule
\end{tabular*}
\end{table*}

\begin{table*}[t]
\centering
\footnotesize
\caption{Fit diagnostics moved from the main-text CE table. The CI column reports the 95\% bootstrap forecast interval for model and joint scaling, and the 95\% bootstrap interval at the last held-out checkpoint for data scaling across all six model sizes. Rows report training-set $R^2$/adjusted $R^2$, and Sig. gives the corresponding fit-significance code.}
\label{tab:appendix_ce_fit_stats}
\setlength{\tabcolsep}{3.2pt}
\renewcommand{\arraystretch}{1.1}
\begin{tabular*}{\textwidth}{@{\extracolsep{\fill}} llccc @{}}
\toprule
Scale & Objective & 95\% CI & Train $R^2$/Adj. & Sig. \\
\midrule
\multirow{3}{*}{Model}
  & Pointwise & \shortstack{$400M$: [3.961, 3.972]\\$1B$: [3.955, 3.969]} & 0.986/0.985 & *** \\
  & Pairwise  & \shortstack{$400M$: [2.034, 2.661]\\$1B$: [1.644, 2.346]} & 0.638/0.619 & *** \\
  & Listwise  & \shortstack{$400M$: [3.716, 3.777]\\$1B$: [3.683, 3.774]} & 0.847/0.839 & *** \\
\midrule
\multirow{3}{*}{Data (17M)}
  & Pointwise & [4.071, 4.082] & 0.849/0.843 & *** \\
  & Pairwise  & [3.839, 3.875] & 0.895/0.890 & *** \\
  & Listwise  & [3.905, 3.932] & 0.883/0.878 & *** \\
\midrule
\multirow{3}{*}{Data (32M)}
  & Pointwise & [4.015, 4.025] & 0.908/0.904 & *** \\
  & Pairwise  & [3.737, 3.785] & 0.882/0.876 & *** \\
  & Listwise  & [3.841, 3.871] & 0.845/0.837 & *** \\
\midrule
\multirow{3}{*}{Data (68M)}
  & Pointwise & [3.977, 3.998] & 0.848/0.841 & *** \\
  & Pairwise  & [3.709, 3.753] & 0.864/0.857 & *** \\
  & Listwise  & [3.783, 3.829] & 0.713/0.699 & *** \\
\midrule
\multirow{3}{*}{Data (150M)}
  & Pointwise & [3.961, 3.983] & 0.869/0.863 & *** \\
  & Pairwise  & [2.094, 2.282] & 0.903/0.898 & *** \\
  & Listwise  & [3.755, 3.795] & 0.779/0.768 & *** \\
\midrule
\multirow{3}{*}{Data (400M)}
  & Pointwise & [3.946, 3.965] & 0.823/0.815 & *** \\
  & Pairwise  & [2.651, 3.041] & 0.122/0.100 & ** \\
  & Listwise  & [3.728, 3.770] & 0.674/0.658 & *** \\
\midrule
\multirow{3}{*}{Data (1B)}
  & Pointwise & [3.952, 3.976] & 0.502/0.479 & *** \\
  & Pairwise  & [2.046, 2.127] & 0.946/0.944 & *** \\
  & Listwise  & [3.707, 3.738] & 0.493/0.469 & *** \\
\midrule
\multirow{3}{*}{Joint}
  & Pointwise & \shortstack{$400M$: [3.946, 3.970]\\$1B$: [3.929, 3.967]} & 0.908/0.906 & *** \\
  & Pairwise  & \shortstack{$400M$: [2.265, 2.547]\\$1B$: [1.944, 2.236]} & 0.619/0.611 & *** \\
  & Listwise  & \shortstack{$400M$: [3.719, 3.757]\\$1B$: [3.691, 3.749]} & 0.885/0.882 & *** \\
\bottomrule
\end{tabular*}
\end{table*}

\begin{table*}[t]
\centering
\scriptsize
\caption{Multiplicative joint-scaling diagnostics for NDCG@10. Train $R^2$/adjusted $R^2$ are computed on the 17M--150M grid. Error columns report Test RMSE / Test MAE / mean bias across all held-out checkpoints of each unseen model. The CI column gives the 95\% bootstrap interval at the last checkpoint, and Coef. reports $(a,c,e)$ from $a+bN^cD^e$.}
\label{tab:appendix_mult_joint_ndcg}
\resizebox{\textwidth}{!}{%
\begin{tabular}{lcccccc}
\toprule
Objective & Train $R^2$/Adj. & Error (400M) & Error (1B) & 95\% CI & Sig. & Coef. \\
\midrule
Pointwise & 0.856/0.854 & 0.0615/0.0526/$-0.0526$ & 0.0939/0.0898/$-0.0898$ & \shortstack{$400M$: [0.3973, 0.4297]\\$1B$: [0.4259, 0.4805]} & *** & $(1.000,-0.0946,-0.0901)$ \\
Pairwise  & 0.850/0.848 & 0.0360/0.0248/$-0.0189$ & 0.0419/0.0349/$-0.0332$ & \shortstack{$400M$: [0.3848, 0.4203]\\$1B$: [0.3971, 0.4467]} & *** & $(0.5727,-0.1514,-0.2303)$ \\
Listwise  & 0.893/0.891 & 0.0383/0.0263/$-0.0257$ & 0.0313/0.0304/$-0.0298$ & \shortstack{$400M$: [0.3640, 0.4062]\\$1B$: [0.3727, 0.4312]} & *** & $(0.4732,-0.2958,-0.2404)$ \\
\bottomrule
\end{tabular}}
\end{table*}

\begin{table*}[t]
\centering
\scriptsize
\caption{Multiplicative joint-scaling diagnostics for CE. Train $R^2$/adjusted $R^2$ are computed on the 17M--150M grid. Error columns report Test RMSE / Test MAE / mean bias across all held-out checkpoints of each unseen model. The CI column gives the 95\% bootstrap interval at the last checkpoint, and Coef. reports $(a,c,e)$ from $a+bN^cD^e$.}
\label{tab:appendix_mult_joint_ce}
\resizebox{\textwidth}{!}{%
\begin{tabular}{lcccccc}
\toprule
Objective & Train $R^2$/Adj. & Error (400M) & Error (1B) & 95\% CI & Sig. & Coef. \\
\midrule
Pointwise & 0.877/0.875 & 0.0316/0.0270/0.0232 & 0.0607/0.0550/0.0513 & \shortstack{$400$: [3.9186, 3.9324]\\$1B$: [3.8793, 3.9027]} & *** & $(0.000,-0.0104,-0.0106)$ \\
Pairwise  & 0.606/0.600 & 0.3817/0.3309/$-0.2164$ & 0.2500/0.1016/0.0388 & \shortstack{$400$: [2.2680, 2.5506]\\$1B$: [1.9451, 2.2528]} & ** & $(0.000,-0.1546,-0.0519)$ \\
Listwise  & 0.863/0.861 & 0.0583/0.0460/0.0295 & 0.0540/0.0483/0.0437 & \shortstack{$400$: [3.6902, 3.7441]\\$1B$: [3.6396, 3.7230]} & *** & $(3.4649,-0.1800,-0.1438)$ \\
\bottomrule
\end{tabular}}
\end{table*}

The TREC results show the same qualitative picture with slightly higher variability, especially for MRR. NDCG@10 and MAP remain well behaved across sizes, and the largest models continue to exhibit low held-out error, indicating that the forecasting signal is not specific to MSMARCO-dev. Taken together, these tables show that the 150M data-scaling result in the main paper is representative of the broader trend rather than an isolated case.

\begin{table*}[t]
\centering
\footnotesize
\caption{MSMARCO-dev data-scaling forecasting errors for all six model sizes. Each cell reports Test RMSE / Test MAE on held-out checkpoints from the corresponding model.}
\label{tab:data_scaling_all_models_msmarco}
\setlength{\tabcolsep}{4.5pt}
\renewcommand{\arraystretch}{1.2}
\begin{tabular*}{\textwidth}{@{\extracolsep{\fill}} llcccccc @{}}
\toprule
Metric & Objective & 17M & 32M & 68M & 150M & 400M & 1B \\
\midrule
\multirow{3}{*}{NDCG@10}
  & Pointwise & 0.007/0.007 & 0.023/0.023 & 0.017/0.017 & 0.019/0.019 & 0.007/0.007 & 0.006/0.006 \\
  & Pairwise  & 0.014/0.014 & 0.018/0.018 & 0.017/0.017 & 0.016/0.016 & 0.004/0.004 & 0.006/0.006 \\
  & Listwise  & 0.008/0.007 & 0.013/0.010 & 0.009/0.006 & 0.008/0.006 & 0.006/0.005 & 0.002/0.002 \\
\midrule
\multirow{3}{*}{MAP}
  & Pointwise & 0.006/0.006 & 0.019/0.019 & 0.013/0.013 & 0.016/0.016 & 0.005/0.005 & 0.005/0.005 \\
  & Pairwise  & 0.011/0.011 & 0.016/0.016 & 0.014/0.014 & 0.013/0.013 & 0.004/0.004 & 0.005/0.005 \\
  & Listwise  & 0.007/0.006 & 0.010/0.008 & 0.008/0.006 & 0.007/0.006 & 0.005/0.004 & 0.002/0.002 \\
\midrule
\multirow{3}{*}{MRR}
  & Pointwise & 0.006/0.006 & 0.019/0.019 & 0.013/0.013 & 0.016/0.016 & 0.005/0.005 & 0.005/0.005 \\
  & Pairwise  & 0.011/0.011 & 0.016/0.016 & 0.014/0.014 & 0.013/0.013 & 0.004/0.004 & 0.005/0.005 \\
  & Listwise  & 0.007/0.006 & 0.010/0.008 & 0.008/0.006 & 0.007/0.006 & 0.005/0.004 & 0.002/0.002 \\
\bottomrule
\end{tabular*}
\end{table*}

\begin{table*}[t]
\centering
\footnotesize
\caption{Data-scaling forecasting errors for all six model sizes averaged over TREC DL '19--'23 and DL Hard. Each cell reports mean Test RMSE / mean Test MAE across datasets for the corresponding model size.}
\label{tab:data_scaling_all_models_trec}
\setlength{\tabcolsep}{4.5pt}
\renewcommand{\arraystretch}{1.2}
\begin{tabular*}{\textwidth}{@{\extracolsep{\fill}} llcccccc @{}}
\toprule
Metric & Objective & 17M & 32M & 68M & 150M & 400M & 1B \\
\midrule
\multirow{3}{*}{NDCG@10}
  & Pointwise & 0.010/0.010 & 0.018/0.017 & 0.021/0.021 & 0.018/0.018 & 0.012/0.012 & 0.007/0.007 \\
  & Pairwise  & 0.014/0.013 & 0.016/0.015 & 0.017/0.017 & 0.017/0.017 & 0.005/0.005 & 0.010/0.010 \\
  & Listwise  & 0.011/0.009 & 0.013/0.011 & 0.008/0.007 & 0.011/0.010 & 0.012/0.010 & 0.004/0.004 \\
\midrule
\multirow{3}{*}{MAP}
  & Pointwise & 0.002/0.002 & 0.007/0.007 & 0.006/0.006 & 0.005/0.005 & 0.003/0.003 & 0.003/0.003 \\
  & Pairwise  & 0.004/0.004 & 0.004/0.004 & 0.005/0.005 & 0.007/0.007 & 0.002/0.001 & 0.004/0.004 \\
  & Listwise  & 0.005/0.004 & 0.007/0.006 & 0.005/0.004 & 0.005/0.005 & 0.005/0.004 & 0.002/0.002 \\
\midrule
\multirow{3}{*}{MRR}
  & Pointwise & 0.016/0.015 & 0.019/0.018 & 0.028/0.027 & 0.018/0.017 & 0.017/0.017 & 0.015/0.014 \\
  & Pairwise  & 0.019/0.018 & 0.015/0.015 & 0.021/0.020 & 0.024/0.024 & 0.012/0.011 & 0.012/0.012 \\
  & Listwise  & 0.031/0.028 & 0.026/0.025 & 0.020/0.017 & 0.019/0.017 & 0.017/0.014 & 0.009/0.008 \\
\bottomrule
\end{tabular*}
\end{table*}

\end{document}